\documentclass[twocolumn]{aastex631}

\usepackage{natbib}
\usepackage{epsfig}
\usepackage{graphicx}
\usepackage{subfigure}
\usepackage{float}
\usepackage{amsmath}
\usepackage{color}
\usepackage{amssymb}
\usepackage{apjfonts}

\newcommand{\PMO}{Purple Mountain Observatory, Chinese Academy of Sciences, Nanjing 210033, China}
\newcommand{\GXU}{Guangxi Key Laboratory for Relativistic Astrophysics, Nanning 530004, China}
\newcommand{\USTC}{School of Astronomy and Space Sciences, University of Science and Technology of China, Hefei 230026, China}
\newcommand{\UA}{Department of Physics, The Applied Math Program, and Department of Astronomy, The University of Arizona, Tucson, AZ 85721, USA}

\shortauthors{Wei \& Melia}

\begin{document}

\title{Exploring the Hubble Tension and Spatial Curvature from the Ages of Old Astrophysical Objects}


\correspondingauthor{Jun-Jie Wei}
\email{jjwei@pmo.ac.cn}

\author{Jun-Jie Wei}
\affiliation{\PMO}
\affiliation{\GXU}
\affiliation{\USTC}

\author{Fulvio Melia \thanks{John Woodruff Simpson Fellow.}}
\affiliation{\UA}

\begin{abstract}

We use the age measurements of 114 old astrophysical objects (OAO) in the redshift range $0\lesssim z\lesssim 8$
to explore the Hubble tension. The age of the Universe at any $z$ is inversely
proportional to the Hubble constant, $H_0$, so requiring the Universe to be older than the OAO it contains
at any $z$ will lead to an upper limit on $H_0$. Assuming flat $\Lambda$CDM and setting a Gaussian prior
on the matter density parameter $\Omega_{\rm m}=0.315\pm0.007$ informed by {\it Planck}, we obtain a 95\%
confidence-level upper limit of $H_0<70.6 \rm{~km} \rm{~s}^{-1} \rm{~Mpc}^{-1}$, representing a $2\sigma$
tension with the measurement using the local distance ladder. We find, however, that the inferred upper limit
on $H_{0}$ depends quite sensitively on the prior for $\Omega_{\rm m}$, and the Hubble tension between early-time
and local measurements of $H_{0}$ may be due in part to the inference of both $\Omega_{\rm m}$ and $H_0$
in {\it Planck}, while the local measurement uses only $H_{0}$. The age-redshift data may also be used for
cosmological model comparisons. We find that the $R_{\rm h}=ct$ universe accounts well for the data, with
a reasonable upper limit on $H_{0}$, while Einstein-de Sitter fails to pass the cosmic-age test. Finally,
we present a model-independent estimate of the spatial curvature using the ages of 61 galaxies and the
luminosity distances of 1,048 Pantheon Type Ia supernovae. This analysis suggests that the
geometry of the Universe is marginally consistent with spatial flatness at a confidence level of
$1.6\sigma$, characterized as $\Omega_{k}=0.43^{+0.27}_{-0.27}$.
\end{abstract}

\keywords{Cosmological parameters (339) --- Cosmological models (337) --- Hubble constant (758) --- Galaxy ages (576)
--- Quasars (1319) --- Type Ia supernova (1728)}

\section{Introduction}
\label{sec:Introduction}

Since the first discovery of the expansion of the Universe more than 90 years ago
\citep{1927ASSB...47...49L,1929PNAS...15..168H}, the Hubble constant $H_0$ characterizing
its current expansion rate has been of great interest to astronomers. In the last decade, however,
a significant mismatch has emerged between several early-time and local measurements of $H_0$
(see \citealt{2019NatAs...3..891V,2021CQGra..38o3001D} for recent reviews).  The latest value
of $H_0$ ($=73.2\pm1.3$ km $\rm s^{-1}$ $\rm Mpc^{-1}$; \citealt{2021ApJ...908L...6R})
measured from local Type Ia supernovae (SNe Ia), calibrated by the Cepheid distance ladder, is
in $4.2\sigma$ tension with that inferred from {\it Planck} cosmic microwave background (CMB)
observations interpreted in the context of the standard $\Lambda$CDM model ($H_{0}=67.4\pm0.5$
km $\rm s^{-1}$ $\rm Mpc^{-1}$; \citealt{2020A&A...641A...6P}).  If the unknown systematics
cannot be responsible for the discrepancy, the Hubble tension may imply new physics beyond
$\Lambda$CDM \citep{Melia2020,2020PhRvD.102b3518V}.

In order to resolve the Hubble tension, more independent methods of measuring $H_0$ are required.
For example, the age of the oldest stellar populations in our galaxy can provide an independent
local determination of $H_0$
\citep{2015ApJ...808L..35T,2019JCAP...03..043J,2020JCAP...12..002V,2021PhRvD.103j3533B,2021MNRAS.505.2764B}.
But most age measurements use objects at higher redshifts, which can also constrain other
cosmological parameters (e.g., \citealt{1995Natur.376..399B,1995GReGr..27.1137K,1996Natur.381..581D,
1999ApJ...521L..87A,2000MNRAS.317..893L,2002ApJ...573...37J,2003ApJ...593..622J,2004PhRvD..70l3501C,
2005MNRAS.362.1295F,2005PhRvD..71l3001S,2006PhLB..633..436J,2006PhRvD..73l3530P,2007A&A...467..421D,
2009PhLB..679..423D,2011PhLB..699..239D,2010PhLB..693..509S,2014A&A...561A..44B,2015AJ....150...35W,
2017JCAP...03..028R,2020MNRAS.496..888N,2021ApJ...908...84V}). Very recently, \cite{2021arXiv210510421V}
used the age estimates of high-redshift (up to $z\sim8$) old astrophysical objects (OAO) to derive
an upper limit on $H_0$ by requiring that all OAO at any $z$ must be younger than the age of the
Universe at that redshift. Their study shed some light on the ingredients needed to resolve the
Hubble tension, but to constrain $H_0$ in this manner, one has to assume a background cosmology.
Assuming the validity of $\Lambda$CDM at late times, \cite{2021arXiv210510421V} found a 95\%
confidence-level upper limit of $H_0<73.2$ km $\rm s^{-1}$ $\rm Mpc^{-1}$, marginally consistent
with that measured using the local distance ladder.

Of direct relevance to the principal aim of this paper is the fact that, unlike the cosmic distance
ladder methods that rely on the distances of primary or secondary indicators, the age measurements
of distant objects are independent of each other. The age-redshift relationship of high-$z$ OAO may
therefore provide a whole new perspective on one of the most frontier issues in modern cosmology,
i.e., the spatial curvature of the Universe. Knowing whether the Universe is open, closed, or flat
is crucial for a complete understanding of its evolution and the nature of dark energy
\citep{2006JCAP...12..005I,2007JCAP...08..011C,2007PhRvD..75d3520G,2008JCAP...12..008V}. A significant
deviation from zero spatial curvature would have far-reaching consequences for the inflationary paradigm
and its underlying physics \citep{2005ApJ...633..560E,2006PhRvD..74l3507T,2007ApJ...664..633W,2007PhLB..648....8Z,Melia2020}.

Although a spatially flat universe ($\Omega_{k}=0$) is strongly favored by most of the current
cosmic probes, especially by the {\it Planck} 2018 CMB observations
\citep{2020A&A...641A...6P},\footnote{Some recent studies show that the {\it Planck} 2015 CMB
anisotropy data support a mildly closed Universe (see \citealt{2019Ap&SS.364...82P,2019ApJ...882..158P}
and references therein).} these curvature determinations
are based on the pre-assumption of a particular cosmological model (e.g., $\Lambda$CDM).
But there is a strong degeneracy between the curvature parameter and the dark-energy equation
of state, so it would be better to measure the purely geometric quantity $\Omega_{k}$ from the
data using a model-independent method.  A non-exhaustive set of references attempting to constrain
the value of $\Omega_{k}$ in a model-independent way includes \cite{2006ApJ...637..598B},
\cite{2007JCAP...08..011C}, \cite{2010PhRvD..81h3537S},
\cite{2014ApJ...789L..15L,2016ApJ...833..240L,2018ApJ...854..146L,2018NatCo...9.3833L,2019ApJ...887...36L,
2019ApJ...873...37L,2020MNRAS.491.4960L}, \cite{2014PhRvD..90b3012S}, \cite{2015PhRvL.115j1301R},
\cite{2016PhRvD..93d3517C}, \cite{2016ApJ...828...85Y}, \cite{2017JCAP...01..015L}, \cite{2017ApJ...839...70L},
\cite{2017JCAP...03..028R}, \cite{2017ApJ...847...45W,2020ApJ...898..100W,2021MNRAS.501.5714W},
\cite{2017ApJ...838..160W}, \cite{2017ApJ...834...75X}, \cite{2018JCAP...03..041D}, \cite{2018ApJ...868...29W},
\cite{2018MNRAS.477L.122W}, \cite{2018ApJ...856....3Y}, \cite{2019PhRvL.123w1101C},
\cite{2019PDU....24..274C,2019NatSR...911608C,2021arXiv211200237C},
\cite{2019PhRvD..99h3514L}, \cite{2019ApJ...881..137R}, \cite{2019PhRvD.100b3530Q,2019MNRAS.483.1104Q},
\cite{2020MNRAS.496..708L}, \cite{2020ApJ...897..127W,2020ApJ...888...99W}, \cite{2020ApJ...889..186Z},
\cite{2021MNRAS.506L...1D}, \cite{2021MNRAS.500.2227J}, \cite{2021ApJ...908...84V}, \cite{2021MNRAS.504.3092Y},
and \cite{2021EPJC...81...14Z}.

In this paper, we broaden the base of support for the age measurements of high-$z$ OAO by demonstrating
their usefulness in testing the late-time expansion history and arbitrating the Hubble tension in
different cosmological models. Further, we propose a new model-independent method of determining the
spatial curvature by combining the OAO age-$z$ data with SNe Ia luminosity distances. Using a
polynomial fitting technique, we reconstruct a continuous age-$z$ function representing the
discrete age measurements of OAO without the pre-assumption of any specific cosmological model. The
time-redshift derivative $dt/dz$ can then be approximately obtained by differentiating the age-$z$
function. Then, $dt/dz$ can be transformed into the curvature-dependent luminosity distance
$D_{L}(\Omega_{k};\;z)$ according to the geometric relation derived from the
Friedmann-Lema\^{\i}tre-Robertson-Walker (FLRW) metric. Finally, by carrying out the joint maximum
likelihood analysis on the polynomial fitting and the observed differences between
$D_{L}(\Omega_{k};\;z)$ and the curvature-independent luminosity distances inferred from SNe Ia,
one can simultaneously constrain the curvature parameter $\Omega_{k}$, the polynomial coefficients,
and the SN nuisance parameters in a model-independent way.

The paper is arranged as follows. In \S~\ref{sec:HT}, we briefly describe the age-redshift
test, and then constrain $H_0$ in different cosmological models. In \S~\ref{sec:OmegaK},
we introduce the methodology of measuring $\Omega_{k}$ using OAO age-$z$ and SN Ia data, and
then present the results of our analysis.  We summarize our main conclusions in
\S~\ref{sec:summary}.

\section{Exploration on the Hubble Tension}
\label{sec:HT}
\subsection{The Age-redshift Test}
The theoretical age of the Universe at redshift $z$ is given as
\begin{equation}
t\left(z\right)=\int^{\infty}_{z}\frac{dz'}{\left(1+z'\right)H\left(z',\;\boldsymbol{\theta}\right)}\;,
\label{eq:tU}
\end{equation}
where $H(z,\;\boldsymbol{\theta})$ is the Hubble parameter and $\boldsymbol{\theta}$ stands for the parameters of
the specific cosmological model. All of our analysis in this paper is based on this
expression, which is derived from the FLRW metric. In so doing, we restrict our attention to
the spacetime predicted in the context of general relativity only, though we shall consider possible
model variations consistent with this constraint as prescribed via the choice of stress-energy tensor
in Einstein's equations, which are characterized by the specific model parameters $\theta$.

The age $t_{{\rm obj}, i}$ of an object (e.g., a passive galaxy or quasar) at redshift $z_{i}$
is defined as the difference between the age of the Universe at $z_{i}$ and that when the object
was formed at redshift $z_{f}$. Given that no object was born at the Big Bang
($z_{f}\rightarrow\infty$), the age of the Universe at any redshift should always be greater
than or equal to the age of the oldest astrophysical object (OAO) at the same redshift, i.e.,
$t(z_{i})\geq t_{{\rm obj}, i}$. The difference between $t(z_{i})$ and $t_{{\rm obj}, i}$,
which we denote by $\tau_{\rm inc}$, represents the `incubation' time, or delay factor,
and accounts for the amount of time elapsed since the Big Bang to the formation of the object.

Equation~(\ref{eq:tU}) shows that the age of the Universe at any given redshift is inversely
proportional to the Hubble constant $H_{0}\equiv H(z=0)$. An upper limit on $H_{0}$ can
therefore be obtained by requiring that the Universe be at least as old as the oldest objects
at the corresponding redshifts \citep{2021arXiv210510421V}.  If the value of $H_{0}$ is too
high, then we are in an awkward position that the Universe is younger than the oldest objects
it contains at a given redshift. In Equation~(\ref{eq:tU}), $t(z)$ receives most of its
contribution at late times ($z\leq10$), and is scarcely sensitive to pre-recombination physics.
Therefore, consistency between the high-$z$ upper limits on $H_{0}$ and the local $H_{0}$
measurements offers a stringent test of late-time and/or local new physics, potentially
suggesting the necessity for the latter to operate together with early-time new physics
to completely address the Hubble tension \citep{2020PhRvD.102j3525K,2021CQGra..38r4001K,
2021arXiv210602532K,2021ApJ...912..150D,2021CmPhy...4..123J,2021ApJ...920..159L,
2021PhRvD.104f3524V,2021arXiv210510421V}.

\begin{figure}
\vskip-0.1in
\centerline{\includegraphics[keepaspectratio,clip,width=0.5\textwidth]{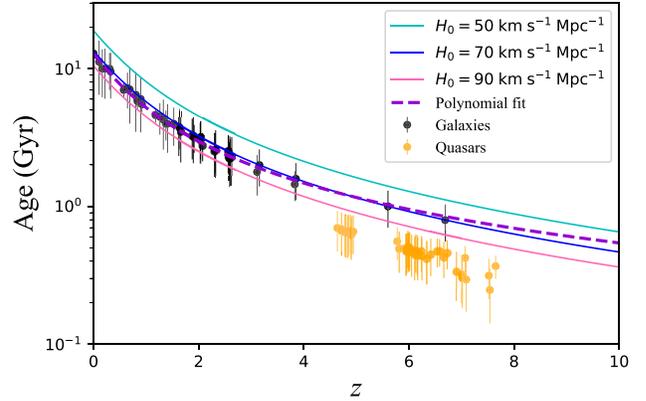}}
\caption{Age-redshift diagram for 114 OAO, including 61 galaxies (black points) and 53 quasars
(orange points).  The solid curves show the age of the Universe as a function of redshift in
flat $\Lambda$CDM with a fixed $\Omega_{\rm m}=0.3$, but an adjustable $H_0$. The violet dashed
curve shows the inferred result when fitting solely the age-redshift data of the 61 galaxies
using a third-order polynomial.}
\label{f1}
\end{figure}

Using a combination of galaxies and high-$z$ quasars, \cite{2021arXiv210510421V} constructed
an age-redshift diagram of OAO up to $z\sim8$. Most of their galaxy data come from the Cosmic
Assembly Near-infrared Deep Extragalactic Legacy Survey (CANDELS) observing program
\citep{2011ApJS..197...35G}, and the remaining galaxy data are from the observations of 32
old passive galaxies in the redshift range $0.117\leq z \leq 1.845$ \citep{2005PhRvD..71l3001S}.
For high-$z$ quasars, they considered the following observations: 7,446 quasars from SDSS DR7
in the range $3\leq z \leq5$ \citep{2011ApJS..194...45S}, 50 quasars detected by the GNIRS
spectrograph in the range $5.5\leq z \leq6.5$ \citep{2019ApJ...873...35S}, 15 quasars detected
by Pan-STARRS1 in the range $6.5\leq z \leq7.0$ \citep{2017ApJ...849...91M}, and 9 of the most
distant quasars ever discovered in the range $7.0\leq z \leq7.642$
\citep{2011Natur.474..616M,2018Natur.553..473B,2018ApJ...869L...9W,2021ApJ...907L...1W,2019ApJ...872L...2M,
2019ApJ...883..183M,2019AJ....157..236Y,2020ApJ...897L..14Y}.
Basically, the ages of the CANDELS galaxies were estimated by fitting the photometric
spectral energy distribution, whereas for the quasars a specific growth model of black hole seeds
developed by \cite{2017ApJ...850L..42P} was adopted.
Applying severe quality cuts to
these observations, and selecting only those objects which are among the oldest ones within
each redshift bin, \cite{2021arXiv210510421V} compiled a final catalog of 114 OAO with reliable
redshift and age measurements, in which 61 OAO are galaxies and the other 53 are quasars. We
adopt this high-$z$ OAO catalog covering the redshift range $0< z < 8$ for our assessment of
the $H_0$ limits. Figure~\ref{f1} shows the age measurements as a function of redshift for
these 114 OAO. In this plot, we also illustrate the dependence of the Universe's age $t(z)$
(estimated using flat $\Lambda$CDM with a fixed matter density $\Omega_{\rm m}=0.3$) on the
value of the Hubble constant $H_0$.

\subsection{Upper Limits on $H_0$}
We are now in position to use the selected 114 age measurements of OAO as a function of redshift
to derive upper limits on $H_0$.  Given the observed data $\mathbf{D}$ (with the OAO ages at
redshifts $z_i$ being $t_{{\rm obj}, i}\pm\sigma_{t_{{\rm obj}, i}}$; see solid points in
Figure~\ref{f1}) and some prior knowledge about the hypothetical models (for which the parameters
are denoted by the vector $\boldsymbol{\theta}$), the posterior probability distributions of the
free parameters can be modeled through the half-Gaussian (log-)likelihood
\citep{2021arXiv210510421V}:
\begin{equation}\label{eq:halfGaussian}
\ln{\mathcal L}\left(\boldsymbol{\theta}\mid\mathbf{D}\right) = -\frac{1}{2}\sum_{i}^{114} \left\lbrace \begin{array}{ll}
\Delta_{i}^{2}\left(\boldsymbol{\theta}\right)/\sigma_{t_{{\rm obj}, i}}^{2}~~~~~{\rm if}~~\Delta_{i}\left(\boldsymbol{\theta}\right)<0\\
0~~~~~~~~~~~~~~~~~~~~~~~{\rm if}~~\Delta_{i}\left(\boldsymbol{\theta}\right)\geq0\;,\\
\end{array} \right.
\end{equation}
where $\Delta_{i}\equiv t\left(\boldsymbol{\theta},\;z_{i}\right)-t_{{\rm obj}, i}$ is defined
as the age of the Universe minus the age of the $i$-th OAO at redshift $z_i$.  The expression
in Equation~(\ref{eq:halfGaussian}) is based on the fact that: $a)$ since the Universe must not
be younger than its oldest inhabitants, parameters for which the Universe is younger than the
OAO (i.e., $\Delta_{i}(\boldsymbol{\theta})<0$) are exponentially unlikely, and this means the
more the Universe is younger than the OAO, the worse the fit; $b)$ parameters for which the
Universe is older than the OAO (i.e., $\Delta_{i}(\boldsymbol{\theta})\geq0$) are equally
likely, and cannot be distinguished solely on the basis of the OAO age.

To calculate model predictions for the age $t(z)$ in Equation~(\ref{eq:tU}), we need an
expression for $H(z,\;\boldsymbol{\theta})$. As the cosmic expansion rate within the context
of specifically selected models is significantly different, it is interesting to examine the
upper limits on $H_{0}$ derived from the OAO ages using different background cosmologies.
Here we discuss how these limits are obtained for $\Lambda$CDM, the Einstein-de Sitter universe,
and the $R_{\rm h}=ct$ universe.

\begin{figure}
\centerline{\includegraphics[keepaspectratio,clip,width=0.5\textwidth]{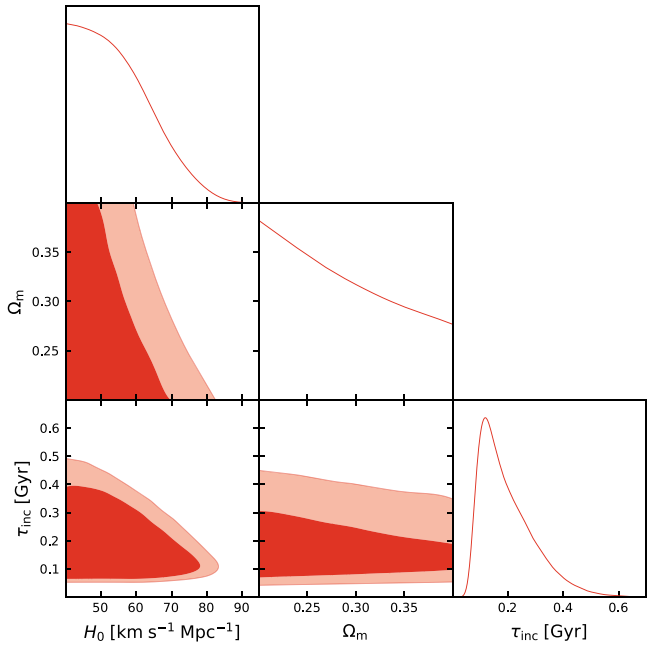}}
\vskip-0.1in
\caption{1D and 2D marginalized posterior distributions with the $1-2\sigma$ contours
for $H_{0}$, $\Omega_{\rm m}$, and the incubation time $\tau_{\rm inc}$, using the
114 age-redshift data shown in Figure~\ref{f1} and the same priors on $H_{0}$,
$\Omega_{\rm m}$, and $\tau_{\rm inc}$ as \cite{2021arXiv210510421V}.}
\label{f2}
\end{figure}

\begin{itemize}
  \item $\Lambda$CDM
\end{itemize}

In flat $\Lambda$CDM, the Hubble parameter is well approximated by
\begin{equation}
H^{{\rm \Lambda CDM}}\left(z,\;\boldsymbol{\theta}\right)=H_{0}\left[\Omega_{\rm m}
\left(1+z\right)^{3}+\Omega_{\Lambda}\right]^{1/2}\;,
\end{equation}
where $\Omega_{\Lambda}=1-\Omega_{\rm m}$ is the cosmological constant energy density.
Note that we ignore the contribution from radiation, which is negligible compared to that
of matter and dark energy in the late-time expansion history. The analysis of the OAO ages
provides a valuable consistency test: if we trust the data, a disagreement between our upper
limit on $H_{0}$ and the value measured from the local distance ladder may indicate new
physics beyond $\Lambda$CDM, at least in the late-time expansion history.

For the basic $\Lambda$CDM model, the free parameters to be constrained are
$\boldsymbol{\theta}=\{H_{0},\;\Omega_{\rm m}\}$. We adopt the Python Markov chain Monte
Carlo (MCMC) module, EMCEE \citep{2013PASP..125..306F}, to explore the posterior probability
distributions of these parameters.  Note that \cite{2021arXiv210510421V} expressed
$\Delta_{i}$ in Equation~(\ref{eq:halfGaussian}) as
$\Delta_{i}\equiv t\left(\boldsymbol{\theta},\;z_{i}\right)-t_{{\rm obj}, i}-\tau_{\rm inc}$,
and modeled the incubation time $\tau_{\rm inc}$ as a prior distribution derived by
\cite{2019JCAP...03..043J}, based on the assumption that the formation redshift $z_f$ for the
oldest observed galaxies is $z_{f}>11$. After marginalizing over $H_{0}$, $\Omega_{\rm m}$,
and $z_f$, this approach yields a prior peaked at $\tau_{\rm inc}\approx0.1-0.15$ Gyr,
which \cite{2021arXiv210510421V} labeled as J19 and adopted its fitting function
provided in Appendix G of \cite{2020JCAP...12..002V}.

For the quasars, \cite{2021arXiv210510421V} fixed
$\tau_{\rm inc}=t(z_{f}=20)$, under the assumption that
they were all seeded at redshift $z_{f}\sim20$. In their baseline analysis,
\cite{2021arXiv210510421V} set flat priors on $H_{0}\in[40,\;100]$ km $\rm s^{-1}$
$\rm Mpc^{-1}$ and $\Omega_{\rm m}\in[0.2,\;0.4]$, the J19 prior on $\tau_{\rm inc}$ for
the galaxies, and fixed $\tau_{\rm inc}=t(z_{f}=20)$ for the quasars. To verify the
reliability of our calculations, we have carried out a parallel analysis with the same
priors on $H_{0}$, $\Omega_{\rm m}$, and $\tau_{\rm inc}$ to ensure that our results
are consistent with each other. Figure~\ref{f2} shows the joint $H_{0}-
\Omega_{\rm m}-\tau_{\rm inc}$ posterior distributions obtained from the baseline
analysis suggested by \cite{2021arXiv210510421V}. Our 95\% confidence-level upper limit on
the reduced Hubble constant $h_{0}\equiv H_{0}/$(100 km $\rm s^{-1}$ $\rm Mpc^{-1}$)
$<0.732$ (all quoted upper limits will hereafter be at the 95\% confidence level)
is the same as that obtained by \cite{2021arXiv210510421V}. Our methodology can thus
reliably incorporate the constraints of \cite{2021arXiv210510421V}, producing results
consistent with their analysis.

As one can see from Equations~(\ref{eq:tU}) and (\ref{eq:halfGaussian}), however, the
inclusion of $\tau_{\rm inc}$ clearly results in a more stringent, less conservative limit
on $H_0$, depending on one's choice of the initial conditions. In addition, the derived
$\tau_{\rm inc}$ distribution from \cite{2019JCAP...03..043J} depends (though only weakly)
on the assumed $\Lambda$CDM cosmology.  In order to be as conservative as possible, and
to provide the most reliable upper limits, we suggest to avoid introducing
$\tau_{\rm inc}$ in Equation~(\ref{eq:halfGaussian}).  For the rest of this section, we
shall therefore begin by conservatively constraining $H_0$ without the inclusion of
this incubation time.

\begin{figure}
\centerline{\includegraphics[keepaspectratio,clip,width=0.45\textwidth]{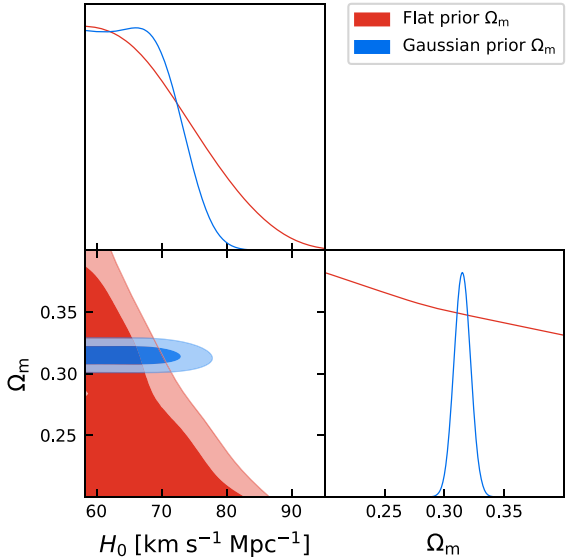}}
\vskip-0.1in
\caption{1D and 2D marginalized posterior distributions with the $1-2\sigma$ contours
for the parameters $H_{0}$ and $\Omega_{\rm m}$ in flat $\Lambda$CDM, using
the 114 age-redshift data shown in Figure~\ref{f1}. Different colored contours correspond to
different priors on $\Omega_{\rm m}$: flat prior $\Omega_{\rm m}\in[0.2,\;0.4]$
(red contours) and Gaussian prior $\Omega_{\rm m}=0.315\pm0.007$ (blue contours).}
\label{f3}
\end{figure}

In our analysis, we choose wide flat priors for $H_{0}\in[0,\;150]$ km $\rm s^{-1}$
$\rm Mpc^{-1}$ and $\Omega_{\rm m}\in[0.2,\;0.4]$. The 1D marginalized posterior
distributions and 2D plots of the $1-2\sigma$ confidence regions for these two parameters,
constrained by the 114 age-redshift data, are displayed in Figure~\ref{f3} (red contours).
These contours show that, whereas $\Omega_{\rm m}$ is not as well constrained, we can set an
upper limit on $H_{0}$, whose 95\% confidence-level value is $h_{0}<0.755$. This is roughly
consistent with its latest local measurement ($h_{0}=0.732\pm0.013$;
\citealt{2021ApJ...908L...6R}). To explore the impact of a $\tau_{\rm inc}$ prior,
\cite{2021arXiv210510421V} also analyzed the data without its inclusion, i.e., by setting
$\tau_{\rm inc}=0$ Gyr, finding in this case that $h_{0}<0.791$, which is somewhat incompatible
with our result ($h_{0}<0.755$). The difference appears to be due to the fact that
\cite{2021arXiv210510421V} set a narrower prior on $h_{0}\in[0.4,\;1]$, while we put
$h_{0}\in[0,\;1.5]$. The relatively low values of $H_{0}$ are equally favored by the
half-Gaussian likelihood (Eqn.~\ref{eq:halfGaussian}).

If one insists on using $\Lambda$CDM as the background cosmology, the version that appears to
be consistent with the majority of observations is spatially flat, with a scaled matter
density $\Omega_{\rm m}\approx0.3$ (e.g., \citealt{2015PhRvD..92l3516A,2018ApJ...859..101S,2020A&A...641A...6P}). Nevertheless, a peculiarity of the often made comparison between the measurements
of $H_0$ at low and high redshifts in this model is that $H_0$ is constrained on its own for the
former, but only in concert with other parameters, particularly $\Omega_{\rm m}$, for the latter.

Thus, to investigate how our results may be affected by the priors for these other concordance
parameters, we sample the limits imposed on $H_0$ using alternate values of the matter density.
First, we adopt the Gaussian prior $\Omega_{\rm m}=0.315\pm0.007$ from {\it Planck}
\citep{2020A&A...641A...6P}.  The resulting constraints are shown as blue contours in
Figure~\ref{f3}.  In this case, the $\Omega_{\rm m}$ posterior unsurprisingly follows its
Gaussian prior, and $h_{0}$ is constrained to be $h_{0}<0.706$, representing a $2\sigma$
tension with its locally measured value.  But it is important to note that it agrees with
the {\it Planck} inference ($h_{0}=0.674\pm0.005$; \citealt{2020A&A...641A...6P}).
This is very interesting because, in this case, the outcomes for both $H_{0}$ and $\Omega_{\rm m}$
are mutually consistent for both {\it Planck} and the OAO age-redshift data.

Next, we explore the impact of an $\Omega_{\rm m}$ prior by fixing its value to be
0.1, 0.3, 0.5, 0.7, and 0.9, respectively. The outcome of each case is presented in
Table~\ref{table1}. One can see that the inferred upper limit on $H_{0}$ does depend
quite significantly on $\Omega_{\rm m}$. That is, some of the impact of adjusting $H_{0}$
for the fits is mitigated by corresponding changes to $\Omega_{\rm m}$. And since the local
measurement of $H_{0}$ does not require $\Omega_{\rm m}$, while {\it Planck} uses both,
the tension between the two measurements may be due in part to the use of
$\Omega_{\rm m}$ in the latter, but not the former.

\begin{table}
\centering \caption{The 95\% confidence-level upper limits on $H_{0}$ with different $\Omega_{\rm m}$ priors}
\begin{tabular}{lccccc}
\hline
\hline
$\Omega_{\rm m}$ (fixed)      &   0.1      &   0.3   &   0.5  &  0.7  &  0.9   \\
$H_{0}$ / [km $\rm s^{-1}$ $\rm Mpc^{-1}$]    &   <112.4     &   <72.2  &   <56.6  &  <47.9  &  <42.4    \\
\hline
\end{tabular}
\label{table1}
\end{table}

\begin{itemize}
  \item The $R_{\rm h}=ct$ universe
\end{itemize}

\begin{figure}
\begin{center}
\includegraphics[width=0.23\textwidth]{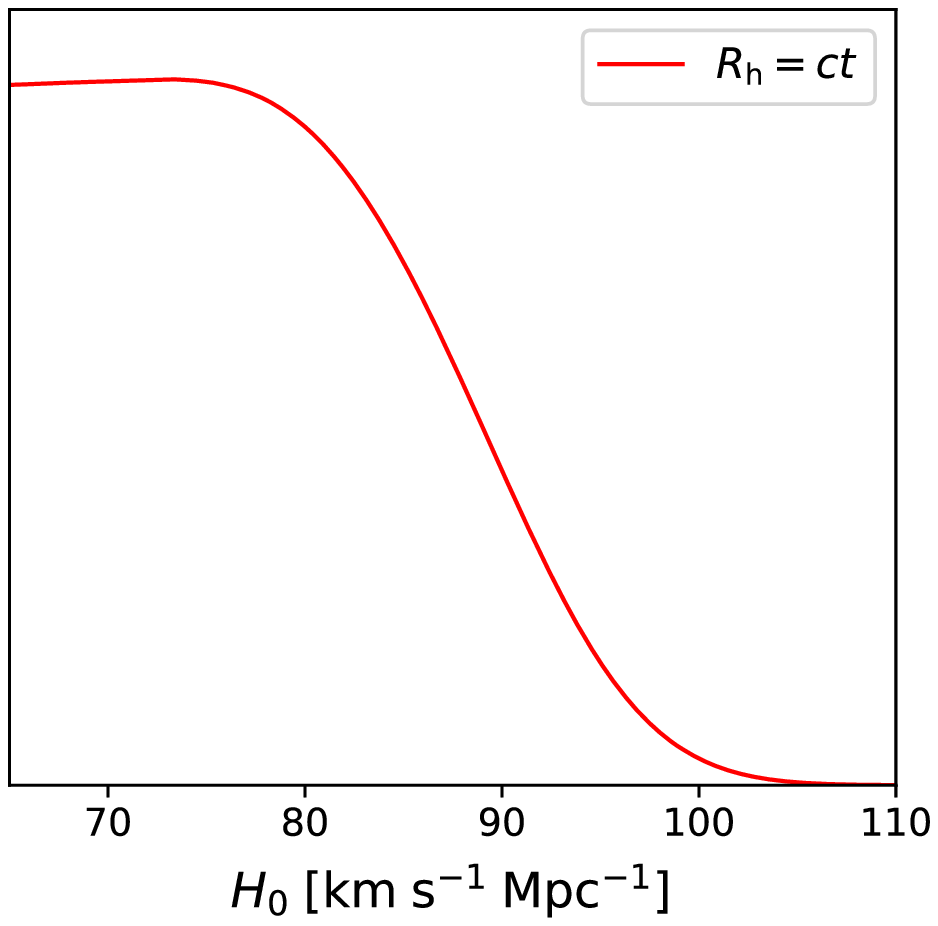}
\includegraphics[width=0.23\textwidth]{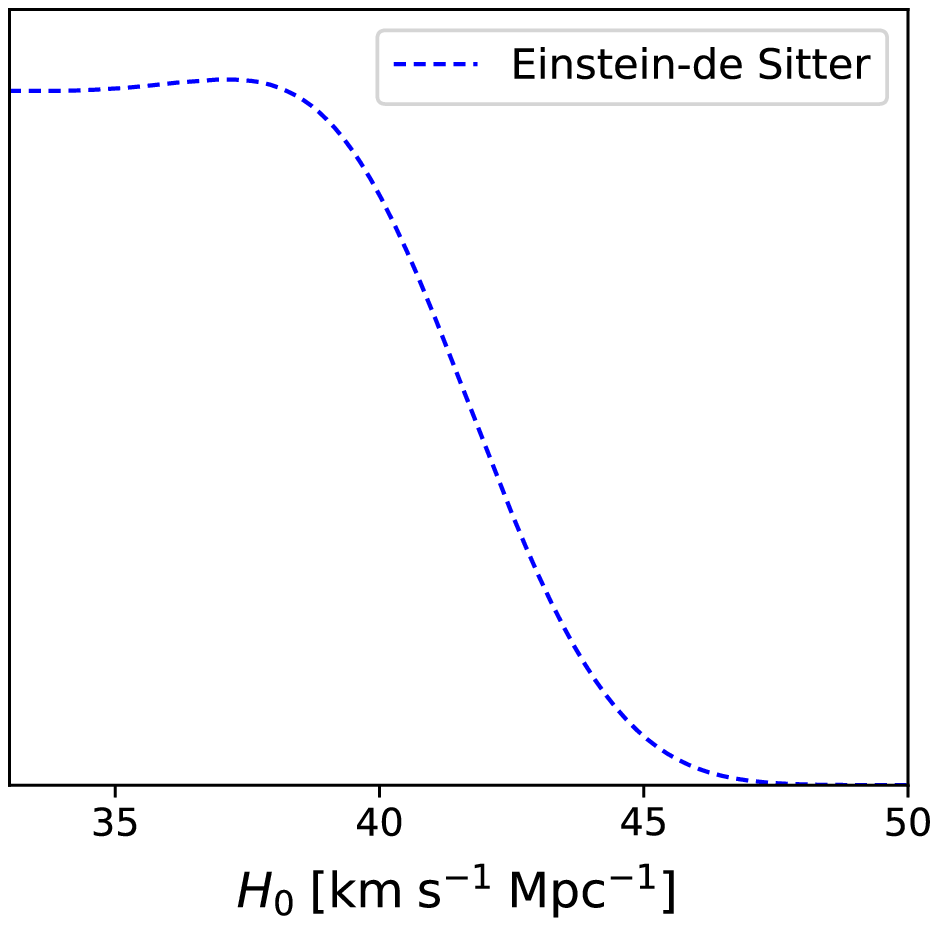}
\vskip-0.1in
\caption{1D posterior distributions of the Hubble constant $H_{0}$ in the $R_{\rm h}=ct$
universe (left panel) and the Einsten-de Sitter universe (right panel), constrained by the
114 age-redshift data.}
\label{f4}
\vskip-0.2in
\end{center}
\end{figure}

The expansion rate in the $R_{\rm h}=ct$ universe \citep{2003eisb.book.....M,2007MNRAS.382.1917M,2013A&A...553A..76M,2012MNRAS.419.2579M,2015AJ....150...35W,Melia2020},
is given as
\begin{equation}
H^{R_{\rm h}=ct}\left(z,\;\boldsymbol{\theta}\right)=H_{0}\left(1+z\right)\;.
\end{equation}
The $R_{\rm h}=ct$ cosmology has only one free parameter, i.e., $\boldsymbol{\theta}=\{H_{0}\}$.
Here we also set a flat prior on $H_{0}\in[0,\;150]$ km $\rm s^{-1}$ $\rm Mpc^{-1}$.
The results of fitting the 114 age-redshift data with this cosmology are shown in the left panel
of Figure~\ref{f4}.  We find an upper limit of $h_{0}<0.861$ at the 95\% confidence-level,
in good agreement with the locally measured $H_{0}$.

\begin{itemize}
  \item Einstein-de Sitter
\end{itemize}

The Einstein-de Sitter universe is characterized by a cosmic fluid containing only matter.
In this model, $H_{0}$ is the sole free parameter, i.e., $\boldsymbol{\theta}=\{H_{0}\}$,
and the Hubble rate is expressed as
\begin{equation}
H^{\rm EdS}\left(z,\;\boldsymbol{\theta}\right)=H_{0}\left(1+z\right)^{3/2}\;.
\end{equation}
With the flat prior on $H_{0}\in[0,\;150]$ km $\rm s^{-1}$ $\rm Mpc^{-1}$, we find that
a low upper limit of $h_{0}<0.401$ is required in order to ensure the Universe is older than
the OAO (see the right panel of Figure~\ref{f4}). The Einsten-de Sitter universe can thus be
safely excluded, given that this inferred upper limit on $H_{0}$ is in $25.5\sigma$ tension with
the locally measured $H_{0}$.

\section{A Cosmology-independent Estimate of the Spatial Curvature}
\label{sec:OmegaK}
In this section, we obtain the curvature-dependent luminosity distance to the OAO, based
on their age measurements, and estimate the spatial curvature constant by comparing it
with the empirically derived distance modulus of SNe Ia.

\subsection{Curvature-dependent distance from the age of OAO}
In the FLRW spacetime, the luminosity distance $D_{L}(z)$ may be written using the first
derivative of the age, $t$, of the Universe,
\begin{eqnarray}
D_{L}\left(z\right)&=& \nonumber\frac{c}{H_{0}} \frac{\left(1+z\right)}{\sqrt{|\Omega_{k}|}}\;\times\\
& &{\rm sinn}\left\{H_{0}\sqrt{|\Omega_{k}|}\int_{z}^{0}\left(1+z'\right)\frac{dt}{dz'}dz'\right\}\;.
\label{eq:DL}
\end{eqnarray}
In this expression, sinn is $\sinh$ when $\Omega_{k}>0$ and $\sin$ when $\Omega_{k}<0$. For
a flat universe with $\Omega_{k}=0$, the right-hand side of this expression simplifies to
the form $(1+z)c$ times the integral. Thus, if one can access the quantity $dt/dz$ at the
required redshift, without pre-assuming a particular cosmological model, one may reconstruct
the line-of-sight comoving distance $D_{C}(z)=c\int_{z}^{0}\left(1+z\right)\frac{dt}{dz'}dz'$
along with the curvature-dependent luminosity distance, $D_{L}(\Omega_{k};\;z)$. The idea
that $dt/dz$ may be obtained from the age-redshift measurements of old objects has been
suggested on various occasions \citep{2002ApJ...573...37J,2017GReGr..49..150J,
2017JCAP...03..028R}.

Since we are primarily interested in the derivative $dt/dz$ and the present age of the Universe
from other observations, and much less so in the incubation time $\tau_{\rm inc}$, we choose to
directly fit the original estimated ages $t_{\rm obj}(z)$ of the OAO. Taking $\tau_{\rm inc}$
to be constant, one may see that $t(z)$ differs from $t_{\rm obj}(z)$ by just a constant. That is,
\begin{equation}
t\left(z\right)=t_{\rm obj}\left(z\right)+\tau_{\rm inc}\;.
\label{eq:tz}
\end{equation}
In principle, we may use the age-redshift data of all 114 OAO up to $z\sim8$ compiled by
\cite{2021arXiv210510421V} to estimate $dt/dz$. However, this catalog includes two different
kinds of source, viz., 61 galaxies and 53 quasars, each of which has its own distinct
incubation time. For this analysis, we therefore only employ the age of 61 galaxies
distributed over the redshift interval $0.001\leq z \leq 6.689$ to estimate $dt/dz$.
The advantage of solely using galaxies is the relative uniformity of the sample.\footnote{
Note, however, that there is no guarantee that all the galaxies constitute a
homogeneous sample either. This analysis should perhaps carried out for each sub-sample
separately. But since the current sub-sample size is admittedly small, we use all the
galaxies for our analysis.} The originally estimated ages, $t_{\rm obj}$, of the 61
galaxies are indicated as a function of redshift by the black points in Figure~\ref{f1}.

In our analysis, we construct the age function $t_{\rm obj}(z)$ in a cosmology-independent
way by fitting a third-order polynomial, with the initial condition
$t_{\rm obj}(z\rightarrow\infty)=0$, to the age-redshift data. To mitigate the convergence
problem that the polynomial fit encounters at high redshifts, we recast the $t_{\rm obj}(z)$
function in the form of the $y$-redshift, defined by the relation $y=z/(1+z)$. In this way,
the age in $z\in[0,\;\infty)$ is mapped into $y\in[0,\;1]$, so that the polynomial fit is
well behaved throughout the redshift range from our local Universe to the Big Bang. This
polynomial is then expressed as
\begin{equation}
t_{\rm obj}\left(y\right)=a_{0}+a_{1}y+a_{2}y^{2}+a_{3}y^{3}\;,
\label{eq:polynomial}
\end{equation}
where $a_{1}$, $a_{2}$, and $a_{3}$ are three free parameters (all in units of Gyr). With
the initial condition $t_{\rm obj}(z\rightarrow\infty)=t_{\rm obj}(y=1)=0$, it is easy to
identify $a_{0}\equiv-a_{1}-a_{2}-a_{3}$.  For $z=0$, Equation~(\ref{eq:tz}) simplifies to
$t_{0}=a_{0}+\tau_{\rm inc}$. Once we have the inferred value of $a_{0}$ and know the
present age of the Universe $t_{0}$, we can also estimate $\tau_{\rm inc}$.

As we assume $\tau_{\rm inc}$ to be constant, we have $\frac{dt_{\rm obj}}{dz}=\frac{dt}{dz}$.
Thus, by differentiating the polynomial (Eqn.~\ref{eq:polynomial}), we obtain
\begin{equation}
\frac{dt}{dz}=\frac{a_{1}}{\left(1+z\right)^{2}}+\frac{2a_{2}z}{\left(1+z\right)^{3}}+\frac{3a_{3}z^{2}}{\left(1+z\right)^{4}}\;.
\label{eq:dtdz}
\end{equation}
Then, the curvature-dependent luminosity distance can be derived by substituting
Equation~(\ref{eq:dtdz}) into (\ref{eq:DL}), i.e.,
\begin{eqnarray}
D_{L}\left(z\right)&=& \nonumber\frac{c}{H_{0}} \frac{\left(1+z\right)}{\sqrt{|\Omega_{k}|}}\;{\rm sinn}\{H_{0}\sqrt{|\Omega_{k}|}\\
& & \times \int_{z}^{0}\left[\frac{a_{1}}{1+z'}+\frac{2a_{2}z'}{\left(1+z'\right)^{2}}+\frac{3a_{3}z'^{2}}{\left(1+z'\right)^{3}} \right]dz'\}\;.
\label{eq:DL_age}
\end{eqnarray}
We can further obtain the reconstructed distance modulus
$\mu_{\rm age}(\Omega_{k},\;a_1,\;a_2,\;a_3;\;z)$ using the age-redshift data:
\begin{equation}
\mu_{\rm age}\left(\Omega_{k},\;a_1,\;a_2,\;a_3;\;z\right)=5\log_{10}\left[\frac{D_{L}\left(z\right)}{\rm Mpc}\right]+25\;.
\label{eq:mu_age}
\end{equation}

\subsection{Distance from observations of SNe Ia}
By comparing the curvature-dependent luminosity distance $D_{L}(\Omega_{k},\;z)$ derived from
the age-redshift data with the empirically-derived luminosity distance (at similar redshifts)
we can obtain a model-independent measurement of $\Omega_{k}$. For the latter, we use the
largest Pantheon SN Ia sample, consisting of 1,048 SNe Ia in the redshift range $0.01<z<2.3$
\citep{2018ApJ...859..101S}.

The observed distance modulus of each SN is given as
\begin{equation}
\mu_{\rm SN}=m_{B}+\alpha x_{1}-\beta {\mathcal C}-M^{\star}_{B}\;,
\label{eq:mu_SN}
\end{equation}
where $m_{B}$ is the observed $B$-band apparent magnitude, $x_{1}$ is the light-curve stretch
factor, and ${\mathcal C}$ is the SN color at maximum brightness. The absolute $B$-band
magnitude $M^{\star}_{B}$ is correlated with the host galaxy mass $M_{\rm stellar}$ via
a simple step function \citep{2014A&A...568A..22B,2018ApJ...859..101S}:
\begin{equation}\label{HSFR}
  M^{\star}_{B} = \left\lbrace \begin{array}{ll} M_{B}+\Delta_{M}~~~~~~~~~{\rm for}~~~M_{\rm stellar}>10^{10}M_{\odot}\\
                                         M_{B}~~~~~~~~~~~~~~~~~~{\rm otherwise}\;, \\
\end{array} \right.
\end{equation}
where $\Delta_{M}$ corresponds to a distance correction based on $M_{\rm stellar}$. Note that
$\alpha$, $\beta$, $M_{B}$, and $\Delta_{M}$ are nuisance parameters that need to be constrained
simultaneously with the cosmological parameters. As such, the derived SN distance is typically
dependent on the chosen cosmology.  To avoid this, \cite{2017ApJ...836...56K} introduced an
approximate method called BEAMS with Bias Corrections (BBC) to correct those expected biases
and simultaneously fit for the SN nuisance parameters. The BBC fit produces a bin-averaged
Hubble diagram of SNe Ia, and then the nuisance parameters $\alpha$ and $\beta$ are constrained
by fitting to a reference cosmological model with fixed values of the matter density
$\Omega_{\rm m}$ and equation-of-sate of dark energy $w$. Within each redshift bin,
the local shape of the Hubble diagram is assumed to be well described by the reference
cosmological model. If there are sufficient redshift bins, the fitted parameters $\alpha$
and $\beta$ will converge to consistent values \citep{2011ApJ...740...72M,2017ApJ...836...56K}.

With the BBC method, \cite{2018ApJ...859..101S} report the corrected apparent magnitudes
$m_{\rm corr}=m_{B}+\alpha x_{1}-\beta {\mathcal C} -\Delta_{M}+\Delta_{B}$ for all the SNe,
where $\Delta_{B}$ is the added distance correction. Given these corrected apparent magnitudes,
we just need to subtract the absolute magnitude $M_{B}$ from $m_{\rm corr}$ to derive the observed
distance moduli:
\begin{equation}
\mu_{\rm SN}=m_{\rm corr}-M_{B}\;.
\label{eq:mu_SNcorr}
\end{equation}

The caveat with this approach, however, is that the format assumes all cosmological models
are nested, which is not true in general. This formulation may be used approximately for
various versions of $\Lambda$CDM, but not for other models, such as $R_{\rm h}=ct$, whose
luminosity distance does not depend on parameters such as $\Omega_{k}$. The caveat here is
that the results we report below pertain specifically to $\Lambda$CDM, not necessarily to
other FLRW models, or models based on alternative theories of gravity.

Even within the context of $\Lambda$CDM, however, there may still be some residual model
dependence, so to test how serious this limitation might be, we take the following approach.
The inferred values of $\alpha$ and $\beta$ in the BBC method are valid only for the
reference model. We therefore consider two different cases: first, the determination
of $\alpha$ and $\beta$ is assumed to be independent of the model, and we directly use
those corrected apparent magnitudes reported by \cite{2018ApJ...859..101S} for our purpose;
second, we carry out a parallel analysis of the uncorrected SN magnitudes by re-constraining
$\alpha$ and $\beta$ as nuisance parameters, and we compare the results.

\subsection{Analysis and results}
We constrain all of the free parameters via a joint analysis involving the galaxy age and SN Ia
data.  The final log-likelihood sampled by the Python MCMC module EMCEE is a sum of the separate
likelihoods of the galaxy ages and SNe Ia:
\begin{equation}
\ln\left({\mathcal L}_{\rm tot}\right) = \ln\left({\mathcal L}_{t_{\rm obj}}\right) + \ln\left(\mathcal{L}_{\rm SN}\right)\;,
\end{equation}
where
\begin{equation}
\ln\left({\mathcal L}_{t_{\rm obj}}\right) = -\frac{1}{2}\sum_{i}^{61}\frac{\left[t_{{\rm obj}, i}^{\rm obs}-t_{\rm obj}^{\rm fit}\left(a_1,\;a_2,\;a_3;\;z_{i}\right)\right]^{2}}{\sigma_{t_{{\rm obj}, i}}^{2}}
\label{eq:Lage}
\end{equation}
and
\begin{equation}
-2 \ln\left(\mathcal{L}_{\rm SN}\right) = \Delta \bf{\hat{\mu}}^{\emph{T}} \cdot \bf{Cov}^{-1} \cdot \Delta \bf{\hat{\mu}}\;.
\label{eq:Lsne}
\end{equation}
In Equation~(\ref{eq:Lage}), $\sigma_{t_{{\rm obj}, i}}$ is the uncertainty of the $i$-th age measurement
$t_{{\rm obj}, i}^{\rm obs}$ and $t_{\rm obj}^{\rm fit}\left(a_1,\;a_2,\;a_3;\;z_{i}\right)$ is obtained from
Equation~(\ref{eq:polynomial}). In Equation~(\ref{eq:Lsne}),
$\Delta \hat{\mu}=\hat{\mu}_{\rm SN}(M_{B};\;z)-\hat{\mu}_{\rm age}(\Omega_{k},\;a_1,\;a_2,\;a_3;\;z)$
is the data vector, defined by the difference between the distance modulus $\mu_{\rm SN}$ of SNe Ia
(Eqn.~\ref{eq:mu_SNcorr}) and the constructed distance modulus $\mu_{\rm age}$ from the galaxy
age-redshift data (Eqn.~\ref{eq:mu_age}), and $\bf{Cov}$ is a full covariance matrix that
contains both statistical and systematic uncertainties of SNe. Note that in the SN likelihood
estimation, there is a degeneracy between $H_0$ and $M_{B}$. We therefore adopt a fiducial
$H_0=70$ km $\rm s^{-1}$ $\rm Mpc^{-1}$ for the sake of constraining $M_{B}$.  In this case,
the free parameters are: the spatial curvature parameter $\Omega_{k}$, the three polynomial
coefficients ($a_1$, $a_2$, $a_3$), and the SN absolute magnitude $M_{B}$.

\begin{figure*}
\vskip-0.1in
\centerline{\includegraphics[keepaspectratio,clip,width=0.8\textwidth]{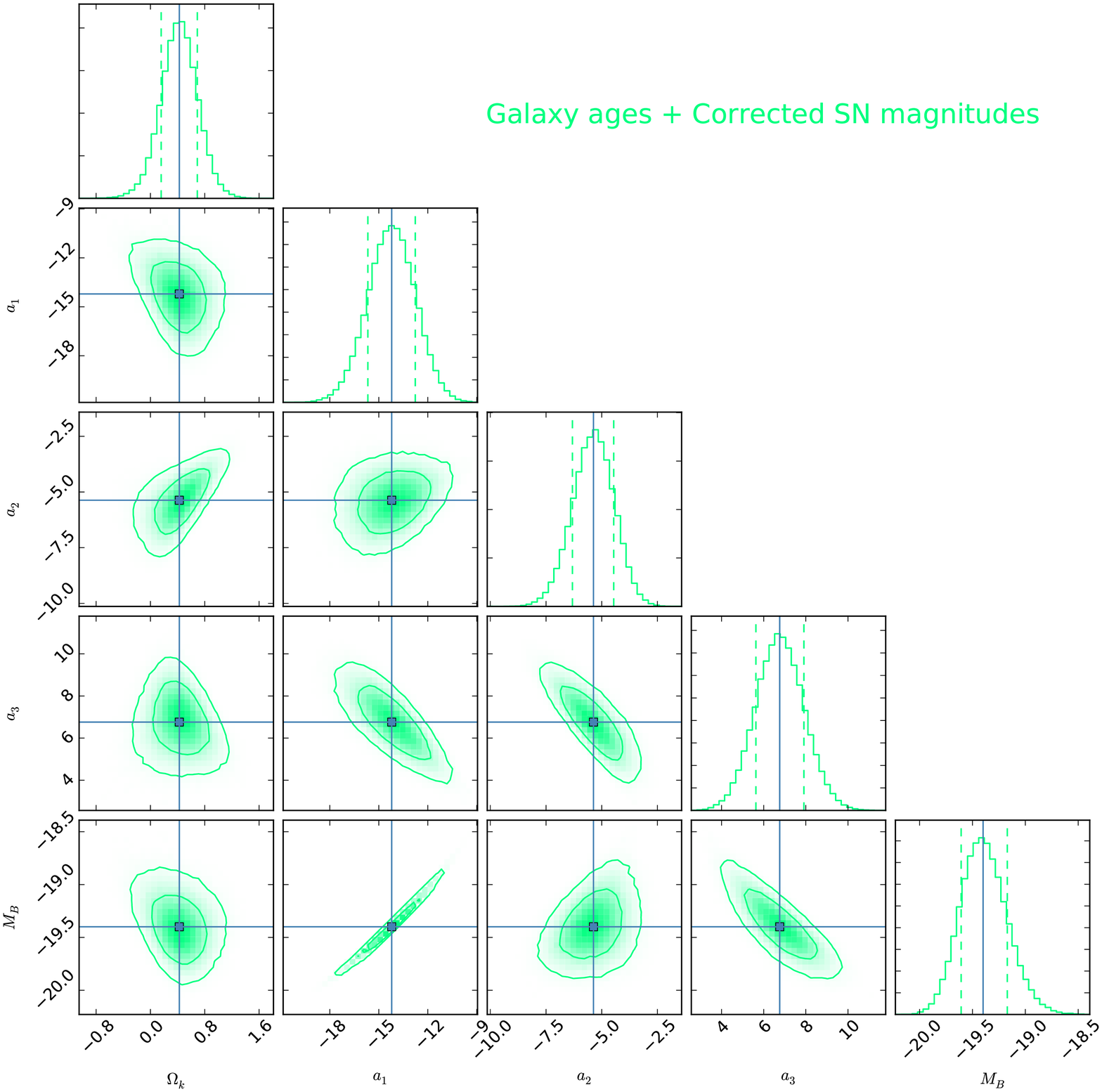}}
\vskip-0.1in
\caption{1D and 2D marginalized posterior distributions with the $1-2\sigma$ contours
for the cosmic curvature $\Omega_{k}$, the polynomial coefficients ($a_1$, $a_2$, $a_3$),
and the SN absolute magnitude $M_{B}$, based on the joint analysis of the galaxy age and corrected SN magnitude data.
The vertical solid lines represent the medium parameter values, whereas the vertical dashed lines
indicate $\pm1\sigma$ deviations from their respective means.}
\label{f5}
\end{figure*}

\begin{table*}
\centering \caption{Constraints on all parameters with different choices of data}
\begin{tabular}{lccccccccc}
\hline
\hline
 Data &  $\Omega_{k}$  &  $a_1$  &  $a_2$   &  $a_3$   &  $M_B$  &   $\alpha$  &   $\beta$   &  $\Delta_{M}$   &  $\sigma_{\rm int}$  \\
      &                &  (Gyr)  &  (Gyr)   &  (Gyr)   &         &             &             &                 &                      \\
\hline
galaxy + corrected SN & $0.43^{+0.27}_{-0.27}$ & $-14.21^{+1.44}_{-1.46}$ & $-5.37^{+0.91}_{-0.95}$ & $6.76^{+1.15}_{-1.13}$  & $-19.40^{+0.23}_{-0.21}$ & -- & -- & -- & -- \\
galaxy + uncorrected SN & $0.59^{+0.18}_{-0.17}$ & $-15.28^{+1.53}_{-1.54}$ & $-3.78^{+0.71}_{-0.74}$ & $6.02^{+1.06}_{-1.05}$  & $-19.48^{+0.23}_{-0.21}$ & $0.132^{+0.005}_{-0.005}$ & $2.595^{+0.057}_{-0.056}$ & $0.052^{+0.009}_{-0.009}$ & $0.079^{+0.006}_{-0.006}$ \\
\hline
\end{tabular}
\label{table2}
\end{table*}

The 1D marginalized posterior distributions and 2D regions with $1-2\sigma$ contours corresponding
to these five free parameters, constrained by the galaxy ages and corrected SN magnitudes, are
presented in Figure~\ref{f5}. These contours show that, at the $1\sigma$ confidence level, the
inferred parameter values are $\Omega_{k}=0.43^{+0.27}_{-0.27}$, $a_1=-14.21^{+1.44}_{-1.46}$,
$a_2=-5.37^{+0.91}_{-0.95}$, $a_3=6.76^{+1.15}_{-1.13}$, and $M_{B}=-19.40^{+0.23}_{-0.21}$. The
corresponding results for the galaxy + corrected SN data are summarized in Table~\ref{table2}.
With this approach, we find that the spatial geometry of the Universe is marginally
consistent with spatial flatness at a $1.6\sigma$ level of confidence.

As noted earlier, our procedure allows us to determine the inferred value of $a_{0}$ along with
the best-fit polynomial coefficients, i.e., $a_{0}\equiv-a_{1}-a_{2}-a_{3}=12.82\pm2.06$ Gyr.
Considering the present age of the Universe as inferred from {\it Planck} in the context
flat $\Lambda$CDM ($t_{0}=13.80\pm0.02$ Gyr; \citealt{2020A&A...641A...6P}), we can further
estimate the incubation time as $\tau_{\rm inc}=t_{0}-a_{0}=0.98\pm2.06$ Gyr.

Next, to investigate how sensitive our results of $\Omega_{k}$ are to the choice of corrected
SN magnitudes provided by the Pantheon team, we also perform a (parallel) comparative analysis
of the galaxy + uncorrected SN data by simultaneously constraining the nuisance parameters along
with $\Omega_{k}$. The likelihood function of SNe now becomes
\begin{equation}
 \mathcal{L}_{\rm SN} = \prod_{i=1}^{1048}\frac{1}{\sqrt{2\pi}\sigma_{{\rm stat},i}}
 \exp\left(-\frac{{\Delta \mu_{i}}^{2}}{2\sigma_{{\rm stat},i}^{2}}\right)\;,
\label{eq:Lsne2}
\end{equation}
where
$\Delta \mu_{i}=\mu_{\rm SN}(\alpha,\;\beta,\;M_{B},\;\Delta_{M};\;z_{i})-\mu_{\rm age}(\Omega_{k},\;a_1,\;a_2,\;a_3;\;z_{i})$
is the difference between the distance modulus $\mu_{\rm SN}$ of SN Ia (Eqn.~\ref{eq:mu_SN})
and the distance modulus $\mu_{\rm age}$ constructed from the galaxy age-redshift data
(Eqn.~\ref{eq:mu_age}), and $\sigma_{{\rm stat},i}$ is the statistical uncertainty of each SN,
given by the expression
\begin{equation}
\begin{split}
\sigma_{{\rm stat},i}^{2}=\sigma^{2}_{m_{B},i}+\alpha^{2}\sigma^{2}_{x_{1},i}+\beta^{2}\sigma^{2}_{\mathcal{C},i}\qquad\qquad\qquad~~~~~\\
+2\alpha C_{m_{B}\,x_{1},\,i}-2\beta C_{m_{B}\,\mathcal{C},\,i}-2\alpha\beta C_{x_{1}\,\mathcal{C},\,i}\qquad~\\
+\sigma^{2}_{\mu-z,i}+\sigma^{2}_{{\rm lens},i}+\sigma^{2}_{\rm int}\;.\qquad\qquad\qquad~~~~~~
\end{split}
\label{eq:sigstat}
\end{equation}
Here, $\sigma_{{m_{B}},i}$, $\sigma_{x_{1},i}$, and $\sigma_{{\mathcal{C}},i}$ stand for the
uncertainties of the peak magnitude and light-curve parameters of the $i$-th~SN, the terms
$C_{m_{B}\,x_{1},\,i},\;C_{m_{B}\,\mathcal{C},\,i}$, and $C_{x_{1}\,\mathcal{C},\,i}$ represent
the covariances among $m_{B},\;x_{1},\;\mathcal{C}$ for the $i$-th~SN, $\sigma_{{\rm lens},i}$
is the uncertainty from stochastic gravitational lensing, and $\sigma_{\rm int}$ is the unknown
intrinsic uncertainty. The dispersion $\sigma_{\mu-z,i}=5\sqrt{\sigma_{z_{\rm pec}}^{2}+
\sigma_{z_{i}}^{2}}/\left(z_{i}\ln10\right)$ accounts for the uncertainty from the peculiar
velocity uncertainty $\sigma_{z_{\rm pec}}$ and redshift measurement uncertainty
$\sigma_{z_{i}}$ in quadrature. We follow \cite{2018ApJ...859..101S} in using
$c\sigma_{z_{\rm pec}}=240$ km $\rm s^{-1}$, as well as $\sigma_{{\rm lens},i}=0.055z_{i}$.

\begin{figure*}
\vskip-0.1in
\centerline{\includegraphics[keepaspectratio,clip,width=1.0\textwidth]{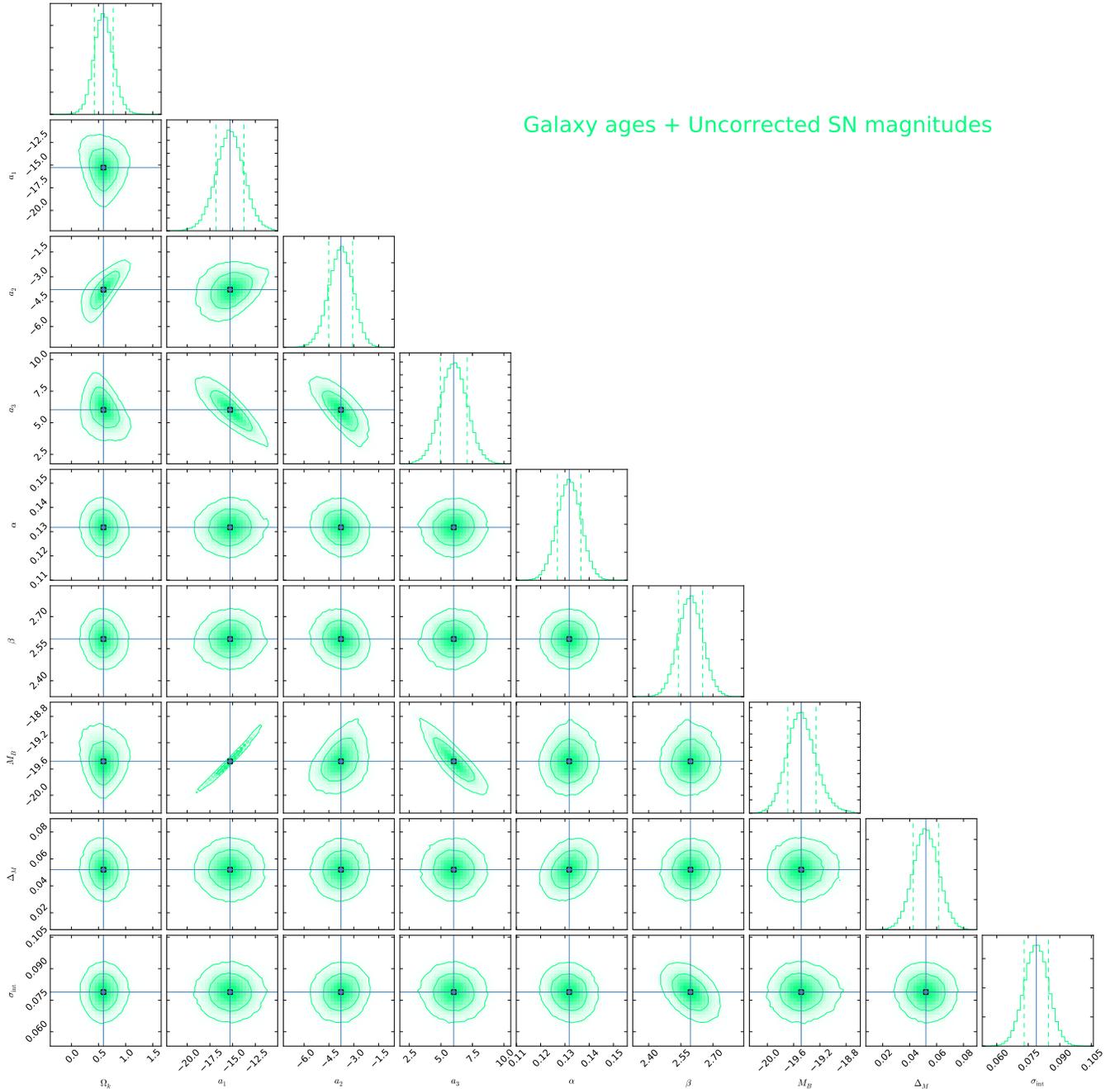}}
\vskip-0.1in
\caption{Same as Figure~\ref{f5}, but now showing the constraints for the parameters
$\Omega_{k}$, $a_1$, $a_2$, $a_3$, $\alpha$, $\beta$, $M_{B}$, $\Delta_{M}$, and
$\sigma_{\rm int}$ based on the joint analysis of the galaxy age and uncorrected SN
magnitude data.}
\label{f6}
\end{figure*}

Only the statistical uncertainties are considered since the six-parameter systematic
covariance matrices ($m_{B}$, $x_1$, $\mathcal C$, $m_{B}\mathcal C$, $x_{1}m_{B}$,
$x_{1}\mathcal C$) are not available in \cite{2018ApJ...859..101S}.  In this case, the
free parameters are the curvature parameter $\Omega_{k}$, the three polynomial coefficients
($a_1$, $a_2$, $a_3$), and the SN nuisance parameters ($\alpha$, $\beta$, $M_{B}$,
$\Delta_{M}$, $\sigma_{\rm int}$).  These nine parameters are constrained to be
$\Omega_{k}=0.59^{+0.18}_{-0.17}$, $a_1=-15.28^{+1.53}_{-1.54}$, $a_2=-3.78^{+0.71}_{-0.74}$,
$a_3=6.02^{+1.06}_{-1.05}$, $\alpha=0.132^{+0.005}_{-0.005}$, $\beta=2.595^{+0.057}_{-0.056}$,
$M_{B}=-19.48^{+0.23}_{-0.21}$, $\Delta_{M}=0.052^{+0.009}_{-0.009}$, and
$\sigma_{\rm int}=0.079^{+0.006}_{-0.006}$, which are displayed in Figure~\ref{f6} and
summarized in Table~\ref{table2}. The comparison between lines 1 and 2 of Table~\ref{table2}
suggests that simply using the corrected SN magnitudes introduces a non-negligible disparity
in the results.  The value of $\Omega_{k}$ inferred from the galaxy + corrected SN data
represents a $0.5\sigma$ tension with that measured from the galaxy + uncorrected SN data.

\section{Summary and Discussion}
\label{sec:summary}
In this work, we have used the age measurements of 114 OAO (including 61 galaxies and 53 quasars)
in the redshift range $0\lesssim z\lesssim 8$ to constrain the late-time cosmic expansion history
and explore the Hubble tension in several cosmological models. Owing to the age of the Universe
at any redshift being inversely proportional to the Hubble constant $H_0$, the requirement that
the Universe be older than the OAO it contains at any redshift provides an upper limit to $H_0$.

Assuming the validity of flat $\Lambda$CDM at late times, and setting wide flat priors on $H_0$
and $\Omega_{\rm m}$, we have obtained $H_0<75.5$ km $\rm s^{-1}$ $\rm Mpc^{-1}$ at the 95\%
confidence level, roughly consistent with local $H_0$ measurements. However, if a Gaussian prior
of $\Omega_{\rm m}=0.315\pm0.007$ informed by {\it Planck} is used, then the 95\% confidence
level upper limit on $H_0$ turned out to be $H_0<70.6$ km $\rm s^{-1}$ $\rm Mpc^{-1}$,
representing a $2\sigma$ tension with the locally measured value. It is compatible with
the {\it Planck} inference, however. This is interesting because, in this scenario, both
$H_{0}$ and $\Omega_{\rm m}$ are mutually consistent for both {\it Planck} and the OAO
age-redshift data. We found that the inferred upper value to $H_{0}$ does depend quite
significantly on $\Omega_{\rm m}$.  Since the local measurement of $H_{0}$ does not require
$\Omega_{\rm m}$, while {\it Planck} uses both, we conclude that the Hubble tension between
the two measurements may be due in part to the use of $\Omega_{\rm m}$ in one case
and not the other.

Besides $\Lambda$CDM, we also discussed how the $H_{0}$ limits may be obtained for
$R_{\rm h}=ct$ and Einstein-de Sitter. The $R_{\rm h}=ct$ universe fits the age-redshift
data with an upper limit of $H_0<86.1$ km $\rm s^{-1}$ $\rm Mpc^{-1}$. By comparison,
the Einstein-de Sitter universe fits the same data with an upper limit of $H_0<40.1$ km
$\rm s^{-1}$ $\rm Mpc^{-1}$. Obviously, Einstein-de Sitter fails to pass the cosmic age test,
because the inferred upper limit to $H_{0}$ in this model represents a $25.5\sigma$ tension
with the locally measured $H_{0}$. Our overall results affirm the idea that cosmic ages are
an extremely valuable probe in the quest towards uncovering the nature of the Hubble tension.

We have also proposed a novel method of estimating the spatial curvature, avoiding possible
biases introduced by the pre-assumption of a specific cosmological model. To perform our
analysis, we have considered the following cosmological data: 61 age measurements of galaxies
and 1,048 SNe Ia from Pantheon compilation.  Based on the geometric relation in the FLRW
metric, we have shown the possibility of obtaining the curvature-dependent luminosity distance
from a best-fit polynomial to the age-redshift data of old objects. By comparing this
curvature-dependent luminosity distance with the empirical luminosity distance inferred
from SNe Ia, we obtained a somewhat model-independent estimate of the curvature parameter
$\Omega_{k}$ based on the parametrization in $\Lambda$CDM.

\cite{2018ApJ...859..101S} applied the BBC method to determine the SN nuisance parameters and
reported the corrected apparent magnitudes for all the Pantheon SNe. Combining the age-redshift
measurements of galaxies with these corrected SN magnitudes, we have placed limits
simultaneously on the cosmic curvature $\Omega_{k}$, the polynomial coefficients
($a_1$, $a_2$, $a_3$), and the SN absolute magnitude $M_{B}$. This analysis suggests that
the curvature parameter is constrained to be $\Omega_{k}=0.43^{+0.27}_{-0.27}$, which
marginally compatible with zero. That is, the spatial geometry of the Universe is marginally
consistent with spatial flatness at the $1.6\sigma$ level.

As the inferred values
of the SN nuisance parameters in the BBC method may depend on the reference cosmological
model, even within the context of $\Lambda$CDM, we also carried out this type of analysis
using the combined galaxy + uncorrected SN data sets by simultaneously constraining the curvature
parameter $\Omega_{k}$, the polynomial coefficients ($a_1$, $a_2$, $a_3$), and the SN nuisance
parameters ($\alpha$, $\beta$, $M_{B}$, $\Delta_{M}$, $\sigma_{\rm int}$). In this case, we
found that the constraint is $\Omega_{k}=0.59^{+0.18}_{-0.17}$.  The value of $\Omega_{k}$
changes slightly, by about $0.5\sigma$, when the SN nuisance parameters are re-constrained
along with the cosmology, implying that simply using the corrected SN magnitudes would
introduce a non-negligible disparity in the results.

Such deviations from zero are not yet compelling enough to initiate a detailed investigation
of their implications. We point out, however, that there are several rather essential
consequences, should such an outcome be realized. First, spatial flatness is assumed to
be an indicator of inflation \citep{1981PhRvD..23..347G}. If the Universe is not spatially flat
afterall, this would cast serious doubt on the possibility that inflation could have
happened. At the very least, it would require major modifications to most of
the inflation potentials proposed thus far. Note, however, that a de Sitter expansion need
not necessarily proceed solely with spatial flatness. As such, several attempts have been
made to create an inflationary scenario with negative spatial curvature, leading to an
an open Universe. This may happen, e.g., in the context of quantum tunnelling-induced
false vacuum decay \citep{1994PhRvD..50.5252R,1994ApJ...432L...5R,1995PhRvD..52.1837R,
1995PhRvD..52.3314B,2012JCAP...06..029K}.

Second, if it turns out that $\Omega_{k}$ is definitely positive, the Universe
must also have net positive energy density \citep{Melia2020}. This would
be very alarming in the context of a quantum-fluctuation origin for the Big Bang, since
it would rule out a `creation from nothing' scenario, in which all the laws of physics,
initial conditions and all the structure appeared as a quantum fluctuation at $t=0$ with
no pre-history. It would at the very least imply a pre-existing vacuum prior to the
expansionary event. Even so, one would then need to contend with the very serious
problem of how a Universe with the known value of Planck's constant and such an enormous
amount of energy could have lived long enough to classicalize and evolve into the
large-scale structure we see today \citep{Melia2020}.

There are good philosphical, if not empirical, reasons for believing that $\Omega_{k}$
must be zero. But we cannot yet make that claim without at least some doubt, certainly
not based on the analysis of the oldest astronomical objects in the Universe that we
have carried out in this paper.

\begin{acknowledgments}
We are grateful to Sunny Vagnozzi for systematically assembling the high-$z$ OAO catalog and
sharing it with us. JJW would like to thank Yan-Mei Han for her infinite patience while part
of this project was carried out at home during the Nanjing lockdown caused by the Covid-19
pandemic. This work is partially supported by the National Natural Science Foundation of China
(grant Nos.~11725314, U1831122, and 12041306), the Youth Innovation Promotion
Association (2017366), the Key Research Program of Frontier Sciences (grant No.
ZDBS-LY-7014) of Chinese Academy of Sciences, the Major Science and Technology
Project of Qinghai Province (2019-ZJ-A10), the China Manned Space Project (CMS-CSST-2021-B11),
and the Guangxi Key Laboratory for Relativistic Astrophysics.
We are also grateful to the anonymous referee for helpful comments.
\end{acknowledgments}


\end{document}